\documentclass[preprint,12pt]{elsarticle}

\usepackage{amssymb}
\usepackage{subcaption}
\usepackage{xcolor}

\newcommand{\red}[1]{#1}

\journal{Journal of Chaos, Solitons \& Fractals}

\begin{document}

\begin{frontmatter}



\title{Towards effective visual analytics on multiplex \red{and multilayer} networks\footnote{This work has been partly supported by the Italian Ministry of Education, Universities and Research FIRB grant RBFR107725. The manuscript has been accepted for publication in Chaos, Solitons and Fractals: the interdisciplinary journal of Nonlinear Science, and Nonequilibrium and Complex Phenomena. The manuscript will undergo copyediting, typesetting, and review of the resulting proof before it is published in its final form. Please note that during the production process errors may be discovered which could affect the content, and all disclaimers that apply to the journal apply to this manuscript.}}
%A definitive version was subsequently published in PUBLICATION, [VOL#, ISSUE#, (DATE)] DOI#"


\author[l]{Luca Rossi}
\address[l]{IT University of Copenhagen, Denmark, lucr@itu.dk}
\author[m]{Matteo Magnani}
\address[m]{Uppsala University, Sweden, matteo.magnani@it.uu.se}

\begin{abstract}
In this article we discuss visualisation strategies for multiplex networks.
Since Moreno's early works on network analysis, visualisation has been one of the main ways to understand networks thanks to its ability to
summarise a complex structure into a single representation highlighting multiple properties of the data. However,
despite the large renewed interest in the analysis of multiplex networks, no study has proposed
specialised visualisation approaches for this context and traditional methods are typically applied instead.
In this paper we initiate a critical and structured discussion of this topic, and claim that the development of
specific visualisation methods for multiplex networks will be one of the main drivers pushing current research results
into daily practice.
\end{abstract}

\begin{keyword}
multiplex \sep visualisation \sep analytics


\end{keyword}

\end{frontmatter}


\section{Introduction}
\label{intro}
Over the last few years multiplex networks have acquired more and more prominence as a promising research direction connecting and complementing many different fields \red{\cite{DBLP:conf/asonam/MagnaniR11,Kivela2014,Boccaletti2014}}. Due to the high level of flexibility of multiplex networks and to a wide range of potential applications the theoretical analysis of these kinds of networks is advancing fast \red{\cite{Brodka2011a,Berlingerio2012,MagnaniSBP2013a,Sole-Ribalta2014}}, showing that single-network theory is not sufficient to handle them. However, when it comes to visualisation traditional schemes are still applied.
Our claim is that shifting from a single-layer to a multiplex perspective also poses new problems concerning network visualisation and about how to handle the richer data that multiplex networks convey. This requires new  approaches challenging some of the traditional dogmas of network visualisation.

%The article is structured as follows. We start with a brief recall of traditional approaches to network data visualisation. Then we move from the limitations of these approaches to a discussion of existing and new data visualisation strategies for multiplexes.

%\subsection{Visualising single networks}

%While goal of this article is not a detailed discussion of all the theoretical and analytical consequences of a multilayer network approach we are, nevertheless, focusing on what are the consequences of such an approach of network data visualisation. 
% OBJECTIVES
Generally speaking the visual representation of network data has two main goals: on the one hand a visual representation can be used as an exploratory tool to obtain relevant insights about the network structure or network properties,
%and about node relations, 
on the other hand it can be used to report the results of a pre-existing analysis in an easily accessible way. In both cases network visualisation involves aspects of information design and geometric representation \cite{brandes2012visualization}.
% FOCUS
These two goals can be applied to two main aspects, leading to completely different visualisations:
%When it comes to network data visualisation we can easily group the existing approaches in two large groups each one aiming at obtaining different results.
we can either focus on the network structure or represent some specific network characteristics or metrics.

\begin{figure}
\begin{subfigure}[b]{.32\textwidth}
\centering\includegraphics[width=\columnwidth]{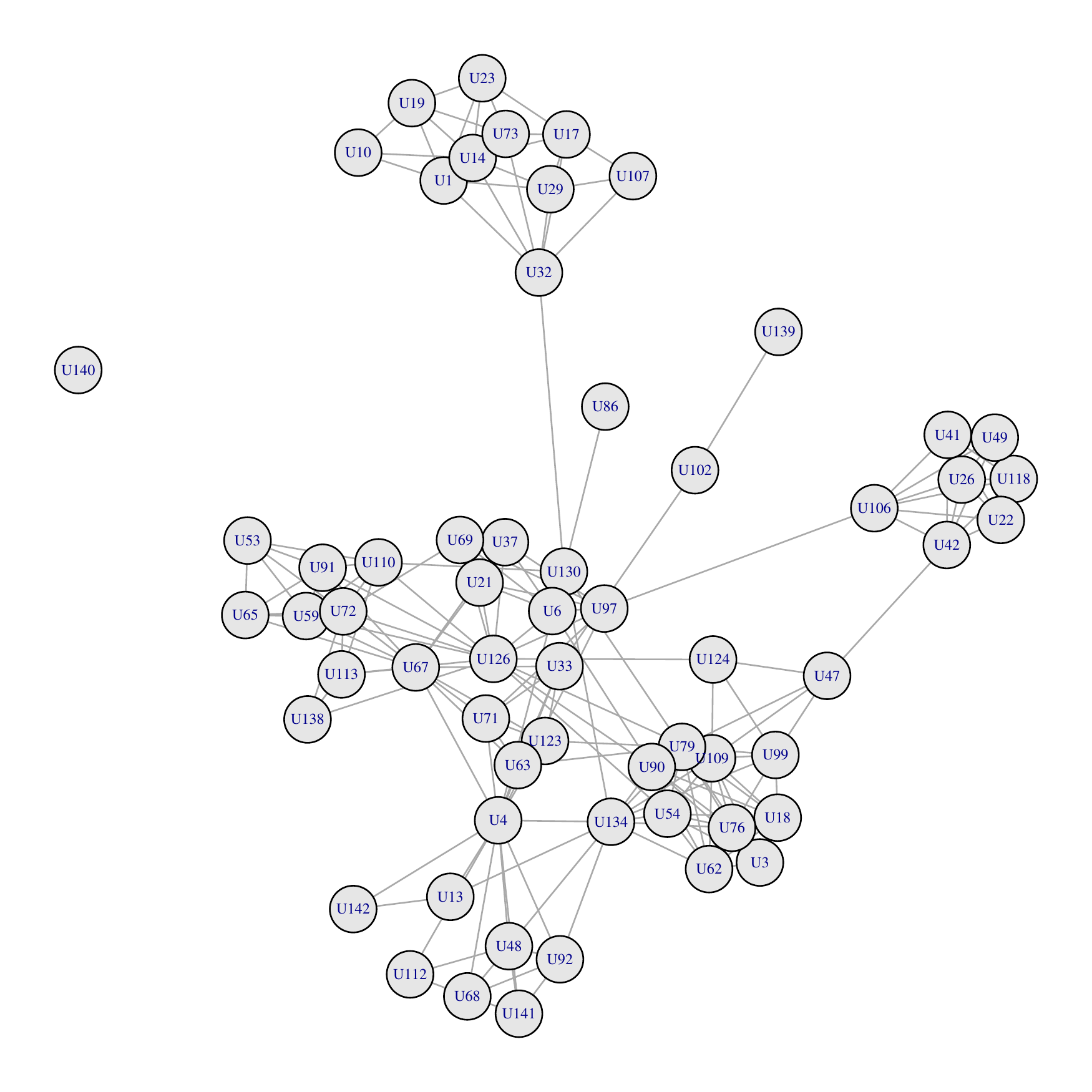}
\caption{}\label{fig:single1}
\end{subfigure}%
\begin{subfigure}[b]{.32\textwidth}
\centering\includegraphics[width=\columnwidth]{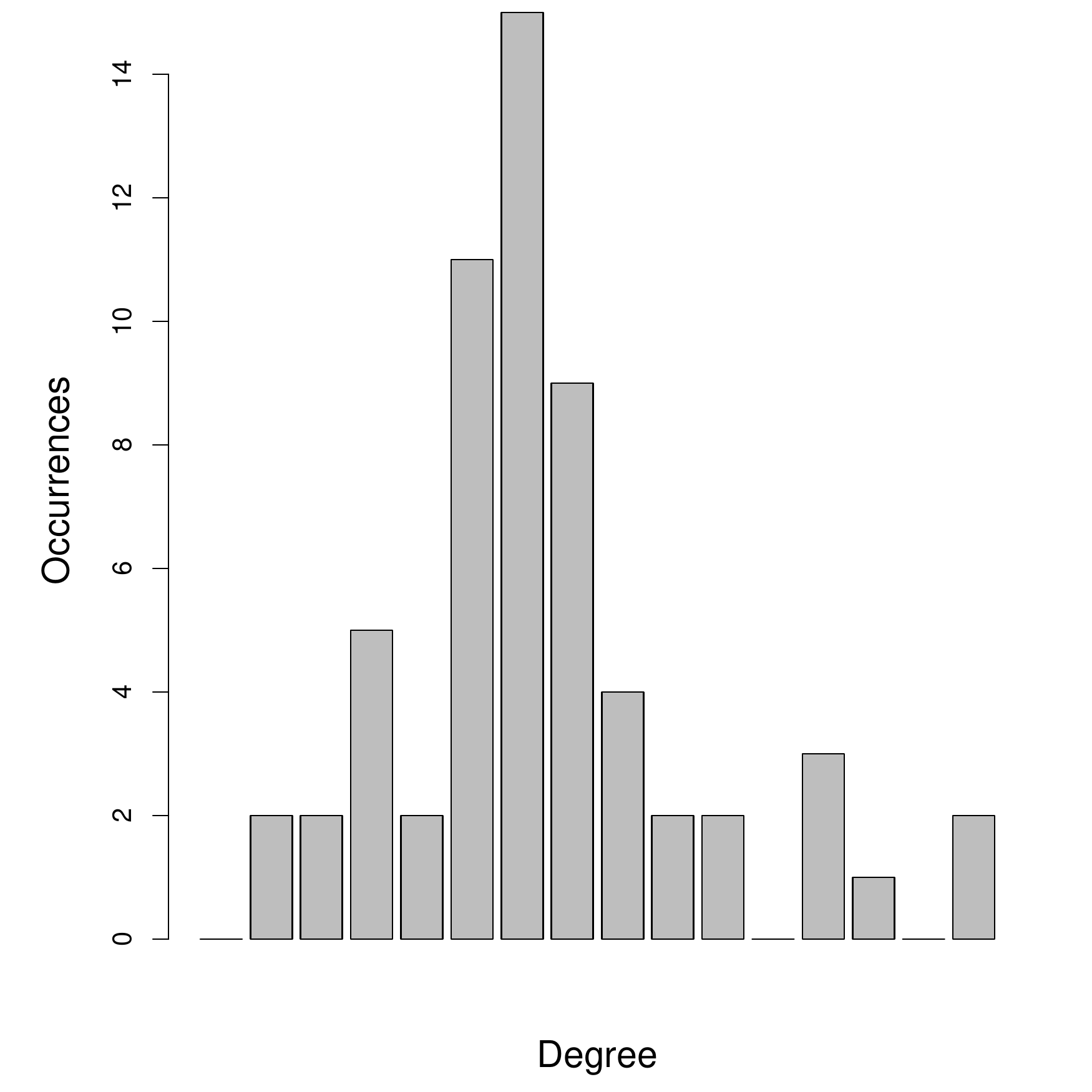}
\caption{}\label{fig:single2}
\end{subfigure}%
\begin{subfigure}[b]{.32\textwidth}
\centering\includegraphics[width=\columnwidth]{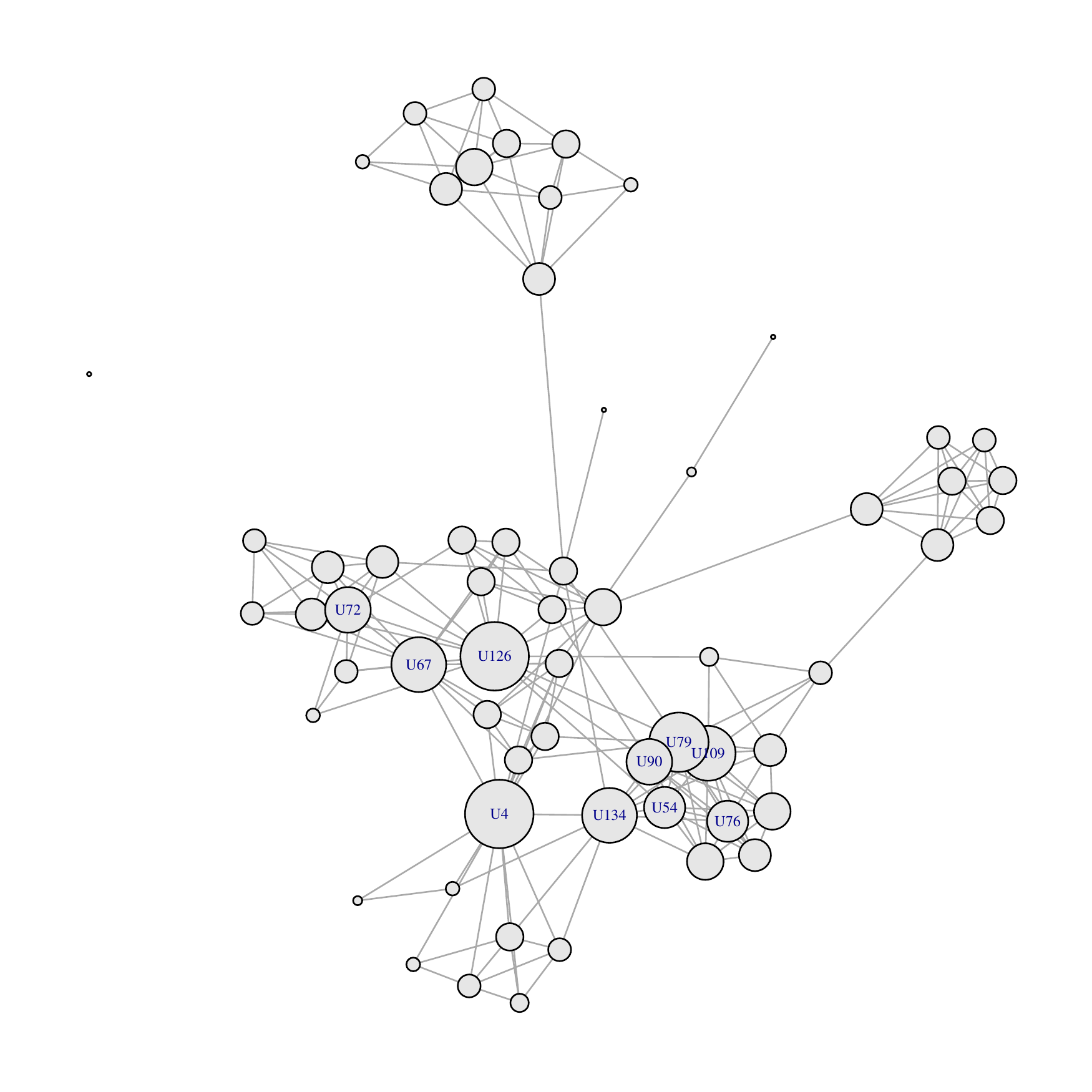}
\caption{}\label{fig:single3}
\end{subfigure}%
\caption{Three typical visualisations of a single network: (a) focus on structure, (b) focus on metrics, (c) augmented visualisation}
\label{fig:single}
\end{figure}

Among the visualisations focusing on the network structure, graphs and their application to social networks known as sociograms \cite{moreno1946sociogram} have seen, so far, the most widespread adoption. A sociogram representing people working at the same department of a University and the relation \emph{having lunch together} is shown in Figure~\ref{fig:single1}.
%In traditional sociograms users are represented by nodes while their relations are represented by the edges connecting them. 
Sociograms put a lot of emphasis on layout, %hile this is undoubtedly a complex computational task for large graphs it is also due to the fact that 
because positional differences are the most accurately perceived graphical attribute \cite{cleveland1984graphical} and also because some network properties like modular structure and node centrality may emerge out of a good layout. This is the case in Figure~\ref{fig:single1}. For a detailed survey on graph visualisation see \cite{DBLP:journals/cgf/LandesbergerKSKWFF11}.

While probably the most appealing, the visualisation of the network structure is not the only way to make sense of a network: sometimes %, when we want to represent or to investigate a specific characteristic of the network, 
it appears to be more convenient to rely on more traditional data visualisation techniques, especially when the network structure does not fit any standard layout or when the network is too large. We can thus describe a network using numbers, either a single value or a distribution, and use traditional visualisations not specifically developed for graphs. A simple example in this context is the study of the degree distribution of a network, which is %While the degree distribution could be in theory visualised also through the sociogram it is 
usually illustrated using a log-log plot (for long tail distributions) or simple histograms like the one about our example \emph{lunch} network shown in Figure~\ref{fig:single2}.
%In this case, while we can make a connection between the degree distribution of a network and its network structure, the focus of the visualisation is not the structure itself but rather a specific metric.

While in this work we focus on the visualisation of the \emph{multiplex network structure}\footnote{For some examples of visualising network metrics please refer to \ref{metrics}.}, it is important to mention the role of network metrics in visual analytics: the two aforementioned options can in fact be merged together, and sociograms can be enriched with information about metrics. In Figure~\ref{fig:single3} the same \emph{lunch}  sociogram is again visualised, but now node sizes represent the node degree. In this way it is possible to communicate information about the network structure (e.g. the presence of clusters) and information regarding single nodes (e.g. degree centrality) at the same time.

Following this idea, this article shows how to use analytical measures to improve the structural visualisation of multiplex networks. However, we will see that the simple application of the same principle to multiplex networks, that is, \emph{augmenting} sociograms with network metrics, does not work as well as in single networks. On the contrary, we explore the possibility of using network metrics to \emph{reduce}, or \emph{simplify} the multiplex. The underlying idea is that multiplex networks carry overabundant information and visualising everything may lead to noise hiding relevant knowledge. But before presenting the details we start our discussion from the existing naive visualisations directly extending single-network approaches.

%\section{Toward a multiplex network visualisation}

\section{From single networks to multiplexes}
\label{network}

\begin{figure}
\begin{subfigure}[b]{.45\textwidth}
\centering\includegraphics[width=\columnwidth]{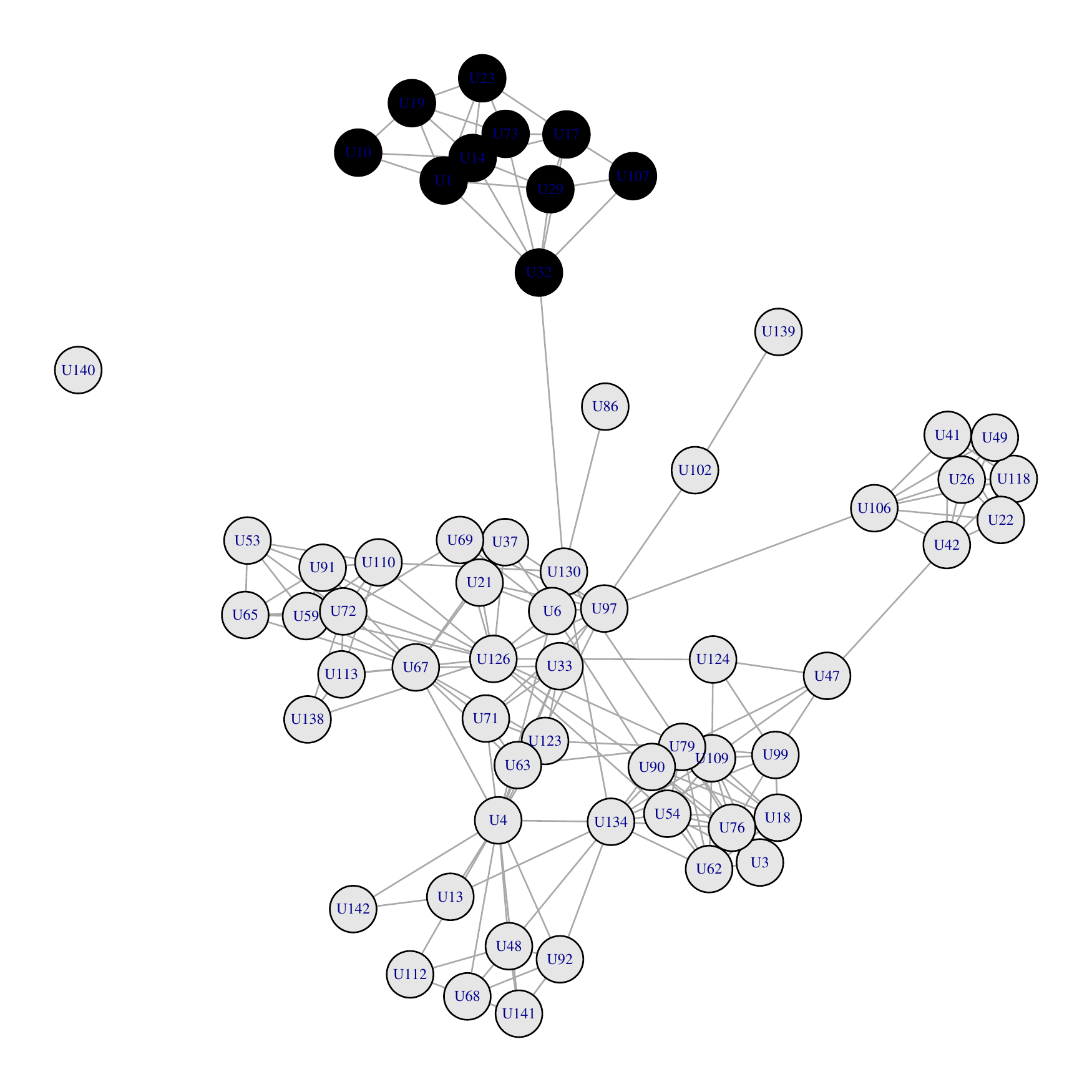}
\caption{A single (lunch) network}\label{Gr-lunch2}
\end{subfigure}%
\begin{subfigure}[b]{.45\textwidth}
\centering\includegraphics[width=\columnwidth]{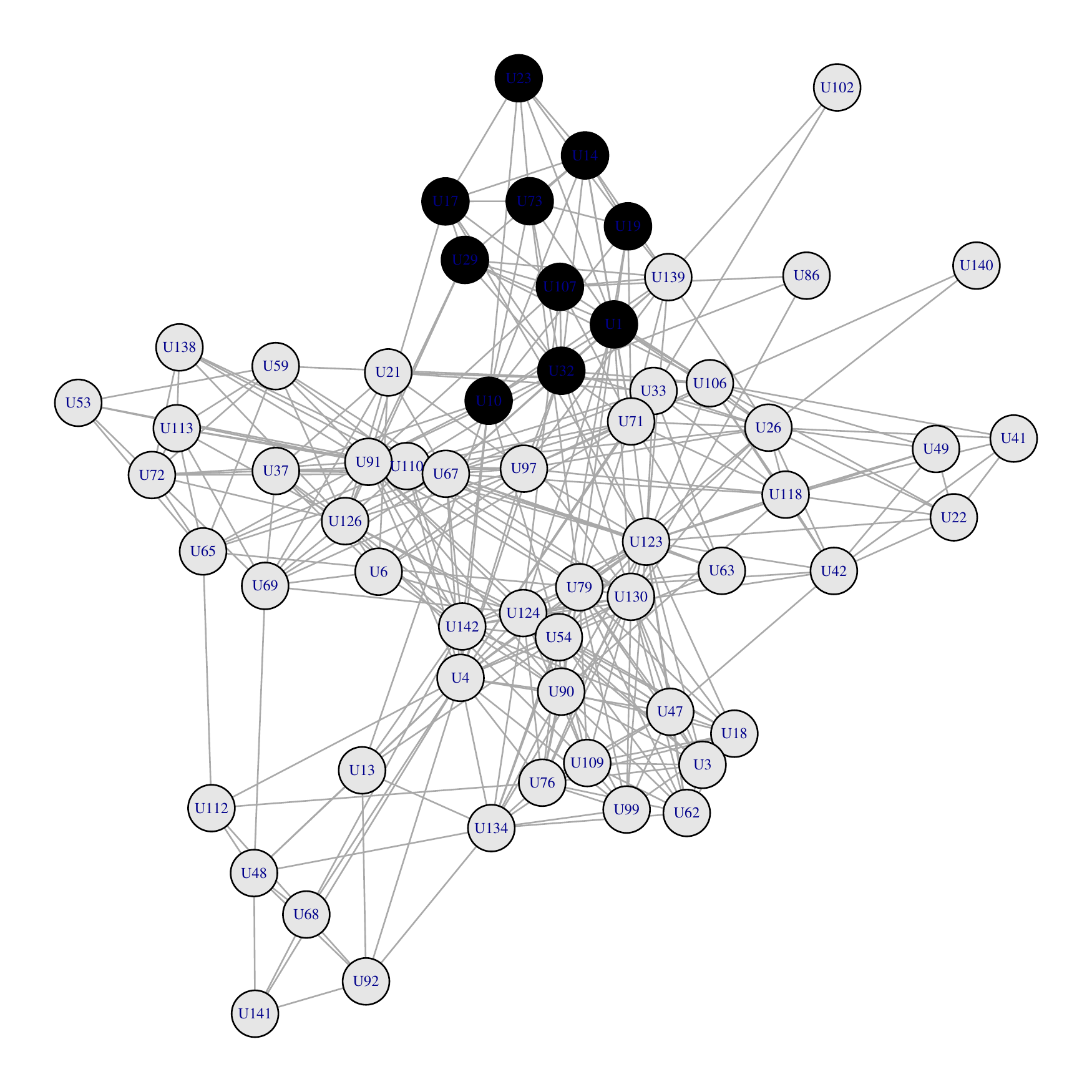}
\caption{The full AUCS multiplex}\label{Gr-appiattita}
\end{subfigure}
\caption{One and five layers of complexity}
\label{fig:flattened}
\end{figure}

When we move into a multiplex network perspective the additional complexity in the added relations makes the \red{choice of an appropriate} layout more challenging. In the following we are going to explore a real-world multiplex network containing five different types of relationships existing among employees of a Computer Science department of a Danish University. The data set counts 61 nodes connected over five different layers: \emph{work, leisure, coauthor, lunch} and \emph{Facebook}. More details about the dataset are provided in \ref{dataset}. Further in the article we will refer to the dataset as AUCS.
%While it is still possible to focus on the representation of (multilayer) structural properties of the network or on some specific metric we will show how the intrinsic complexity and data richness of multilayer networks requires also innovative approaches.
In Figure~\ref{fig:flattened} we can see again the \emph{lunch} layer of the AUCS multiplex network (left) and the whole network with the four additional kinds of relations (right).

\begin{figure}
\centering\includegraphics[width=.6\columnwidth]{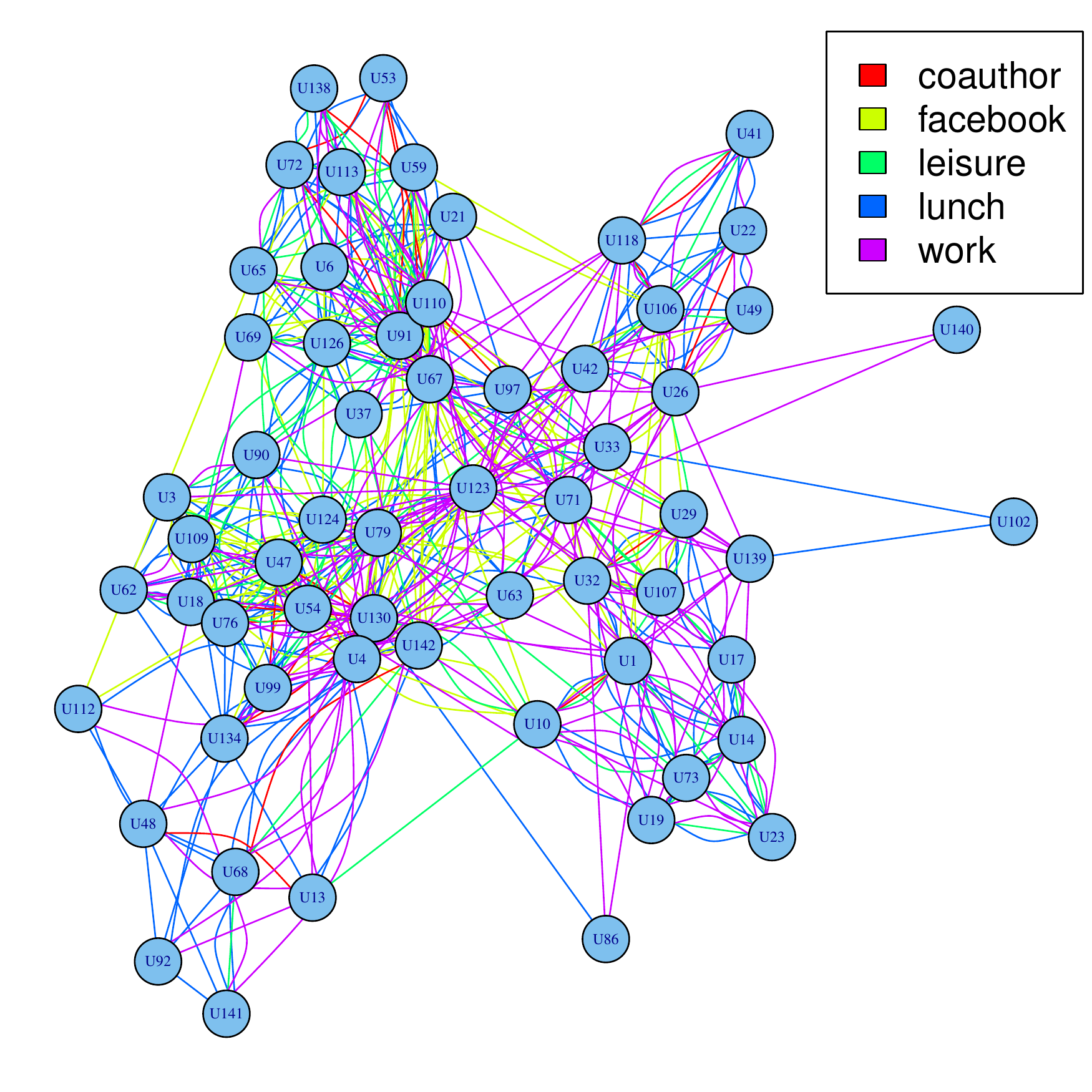}
\caption{Flattened multiplex: colours represent different layers (edge types)}
\label{Gr-multiplex_color}
\end{figure}

Comparing the two visualisations we can see how the clear structure of the \emph{lunch} network becomes more blurred and confused if we take connections from all the layers into consideration. As an example, we have highlighted a clearly visible cluster in the left hand side network using a black node background. The same nodes are also black-marked in the graph with all the multiplex connections, and we can see that the cluster has been partially attracted towards the center of the figure and some of its peripheral nodes are now more connected to other nodes outside the cluster.
%
%At the same time we can easily agree that focusing on a single layer of communication would be a dangerous oversimplification for any real-world problem and defining which are the central layers might not be a trivial task. For example, studying the process of information diffusion within this multiplex network, how can we decide if and to what extend $working together$ is a more relevant layer than $having lunch together$?
% 
%While we can easily agree on the fact that almost every layer provide a unique and valuable information about the whole network, visualising  multiplex structures is probably one of the most interesting and challenging task in contemporary network visualisation. Goal of a good network visualisation should be to provide valuable insight of the inner structure of the network and of the existing clusters or sub structures.  This task 
Network visualisation is already a complex task for single-layer networks when they count a large number of highly interconnected nodes. When it comes to multilayer networks the task is even more challenging and even adopting a 3-dimensional interactive visualisation \cite{de2014multilayer} a multiplex network quickly becomes incomprehensible as soon as it contains a few dozen nodes. 

%Figure~\ref{Gr-appiattita} shows what is probably the simplest strategy to represent a multilayer network within a two-dimensional representation. In  every existing edges on every layer between two nodes has been flattened into a single layer network.
In addition, while the graph in Figure \ref{Gr-appiattita} still shows some structural features, e.g., some denser locations, information about the different layers is completely lost. Therefore, a more typical way to preserve some of the multilayer information is to assign a colour to each layer as in Figure~\ref{Gr-multiplex_color}. However, although fancier, this figure does not add much to the flattened uncoloured case. %Also this edge-coloured graph shows some of the multilayer network structure, but e
Even if some denser regions can still be observed it is almost impossible to understand how those are related to the underlying multiplex structure.
%When many layers overlap the edges create a confused situation with no clearly understandable layout or structures.
In addition, it is very challenging to focus on a specific network \red{layer} within this chaotic overlapping of edges.  

\begin{figure}
\centering\includegraphics[width=\columnwidth]{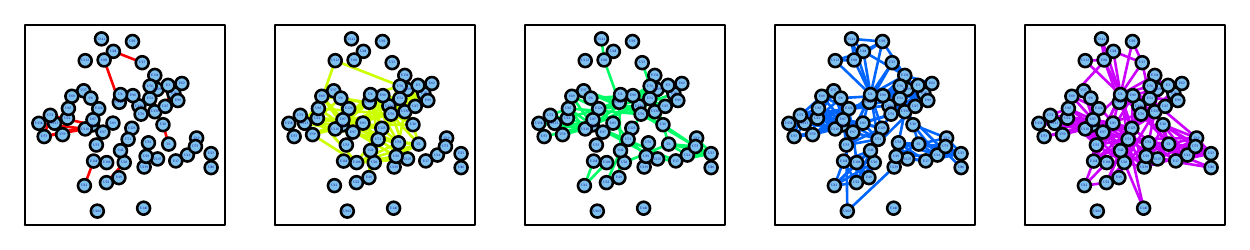}
\caption{Sliced visualisation, same layout}
\label{Gr-multiplex same layout}
\end{figure}

\begin{figure}
\centering\includegraphics[width=\columnwidth]{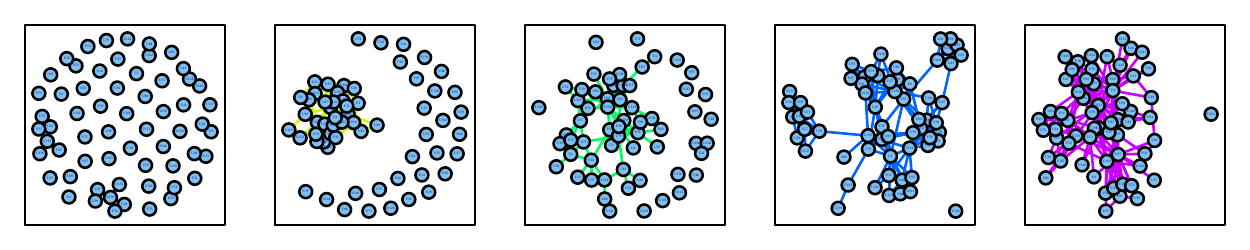}
\caption{Sliced visualisation, independent layouts}
\label{Gr-multiplex slices multi layout}
\end{figure}

Two alternative visualisations concluding our review of typical approaches are shown in Figures~\ref{Gr-multiplex same layout} and \ref{Gr-multiplex slices multi layout}. Both methods just slice the multiplex into its layers. In order to simplify a comparison between the layers, in Figure~\ref{Gr-multiplex same layout} the nodes have been placed using the same layout in each slice\footnote{In this specific case we have computed the common layout on the flattened graph, but any layer can be used to this aim.}. 
%While on the one side this visualisation offers the possibility to compare the various layers observing similarities and differences in the structures it has two major downsides: Separating the multilayer structure into a set of different frames doesn't provide any idea of unity and  
However if different layers contain different connections their internal structure can become invisible.
%; computing the layout positions starting from the aggregated network produce a layout that it not optimal for any layer and might hide, or make less visible, structures that would had been visible with the right layout. You can have a clear perception of 
This is evident e.g. comparing the \emph{lunch} layer in Figure~\ref{fig:flattened} with the same layer visualised in Figure~\ref{Gr-multiplex same layout} (fourth slice, blue edges). If we use a specific layout for every layer as in Figure~\ref{Gr-multiplex slices multi layout} we can better appreciate the structure of each layer but we loose the possibility of detecting structures developing over multiple layers.

\section{Augmented multiplex sociograms}

In a similar way to what happens for single layer networks, analytical measures like the ones defined in \cite{matteoMDMea2013} can be used to increase the information content of the graphs introduced in the previous section.
% Nevertheless there are few unique elements in multiplex networks that requires an extension of classical approaches. While we provide a more detailed description of these in~\ref{metrics} it is important to point out how multiplex networks enable both new dimensions for existing metrics (such as the extension of the degree value in a multiplex context defined in \cite{matteoMDMea2013}) and brand new metrics (such as the interlayer correlation coefficient described in \cite{matteoMDMea2013}). 
The next step in our exploration of visualisation strategies is thus to use some metrics to produce augmented versions of the multiplex sociograms. 

\begin{figure}
\centering\includegraphics[width=.7\columnwidth]{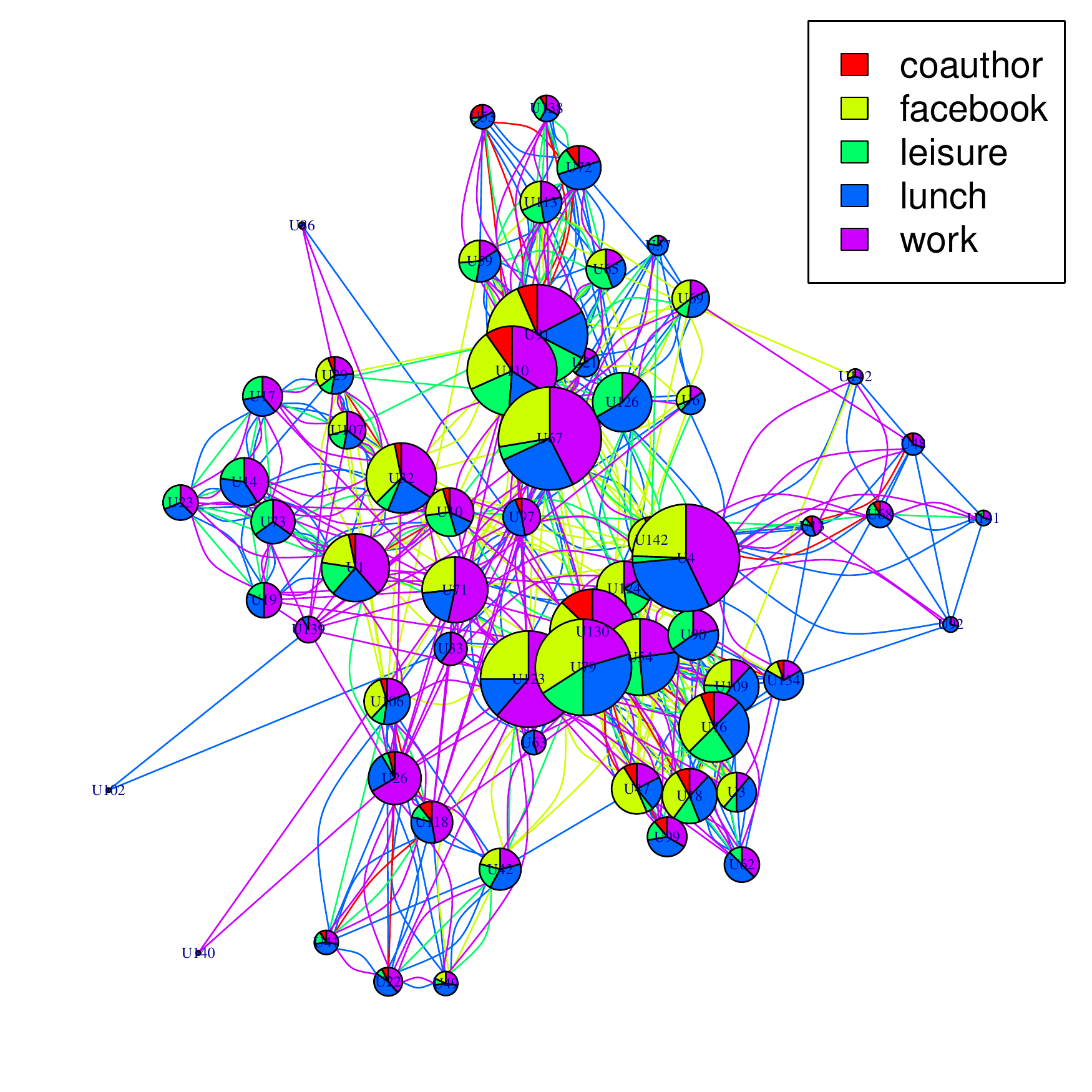}
\caption{Our working example \emph{augmented} with multiplex measures (local degree)}
\label{Gr-multiplex pies}
\end{figure}

Figure~\ref{Gr-multiplex pies} shows a coloured multiplex network  where every node contains information about its degree (node size) and the degree composition on the various layers (pie chart). While this may look like an interesting visualisation it is hard to claim that it provides a clear understanding of the underlying network structure. % or of the node behaviour on the various layers.
%These approaches have pros and cons and should make clear the complexity of visualising network structure data within a multiplex perspective.
%In order to address the complexity in visualising structural information for multiplex networks we introduce two different approaches \emph{local merging} and  \emph{ranked sociogram}: see figures \ref{Gr-parabola} and \ref{merging-relevance}.
The main issue with this visualisation can still be described as \emph{overabundant information}: edges overlap with each others and generate an intricate pattern.

\begin{figure}
\centering\includegraphics[width=.7\columnwidth]{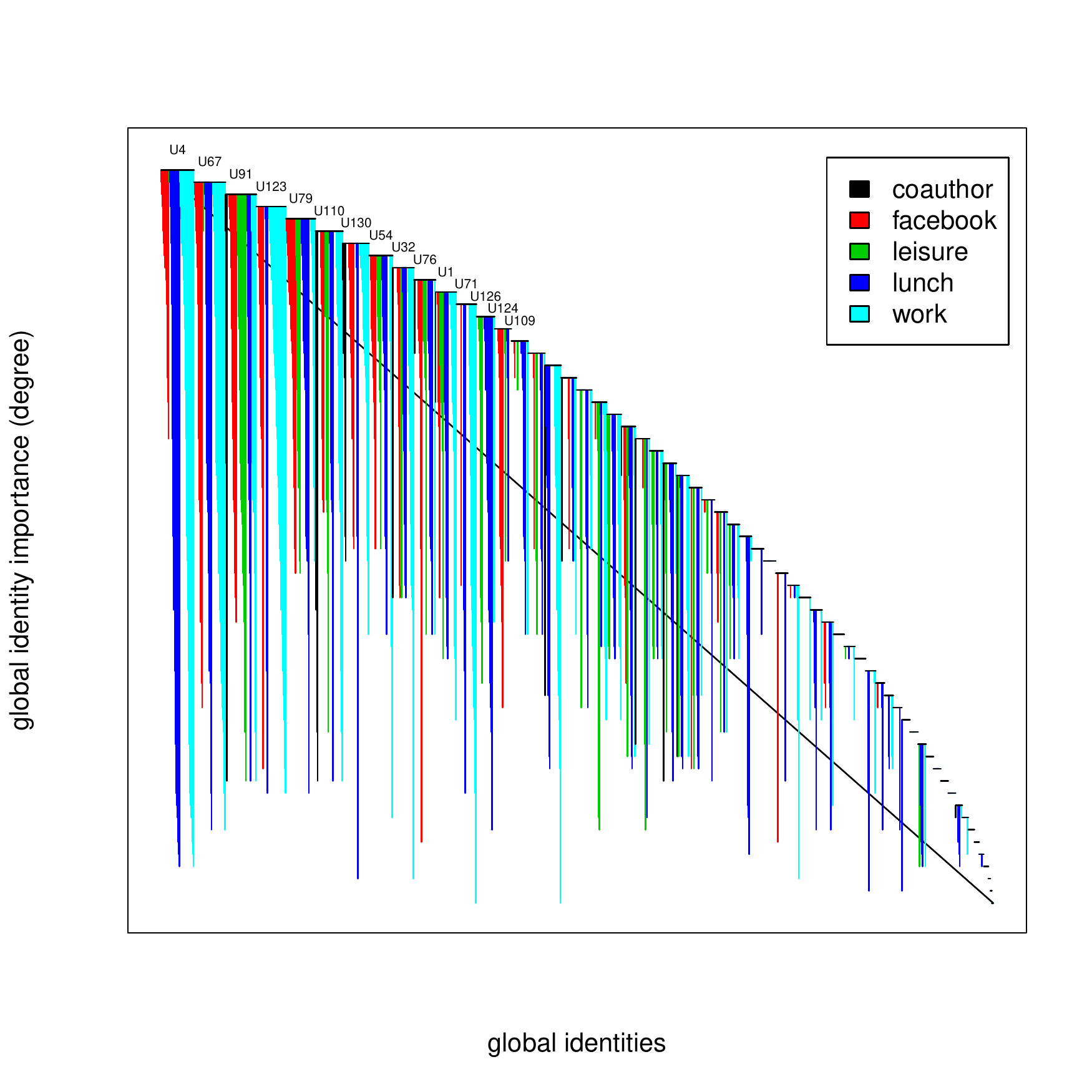}
\caption{Ranked sociogram of our working example}
\label{Gr-parabola}
\end{figure}

In order to remove this intricacy, we have explored alternative visualisations rearranging the edges in a non-standard way so that overlapping is prevented.
%be able to convey the large amount of multiplex information sociograms have to be revised.
An approach following the idea in \cite{citeulike:11570704}, that we call here a \emph{ranked sociogram}, abandons the traditional sociogram  visualisation --- where edges go from one node to the other --- and replaces it with ranking-based positioning of nodes and edges where the ranking is based on a specific metric. In a ranked sociogram nodes are plotted according to a chosen metric on a classical xy plot (Figure~\ref{Gr-parabola} ranks them according to their aggregated degree). Every node has a length on the x axis that is defined by a set of edges connecting the vertical position of node \emph{a} with the vertical position of node \emph{b}. All the edges expand only on the vertical axis and show the distance between the nodes according to the chosen ranking. Every edge is also represented using a different colour according to its corresponding layer. 

This produces a clear perception of how many connections every node has on every single layer and \emph{how far} the connected nodes are in terms of the chosen metric. 
%We could define a \emph{ranked sociogram} as a sociogram where nodes' position is defined by the ranking of the node according to a selected metric while nodes' behaviour on the various layers is expressed through the colour of the edges. 
In Figure~\ref{Gr-parabola} the user with the highest degree (U4) is mainly connected through the \emph{work} layer, the \emph{lunch} layer and the \emph{Facebook} layer while he/she has no connections on the \emph{leisure} and on the \emph{coauthor} layers. It is also interesting to notice how U4 is connected with users with very different degrees (as indicated by the length of the edges) while this is not happening on the \emph{Facebook} layer, spanning a shorter range of contacts. This also gives an insight on the layers' assortativity or dissortativity.
Another interesting element that can be noticed studying Figure~\ref{Gr-parabola} is that between the top 5 users only two (U91 and U79) show a relevant presence on the \emph{leisure} layer\red{, which} appears to be absent for the \red{three other top}  users. More details about the ranked sociogram are available in \ref{ranked}.

\section{Local simplification of multiplex structures}
\label{metricsandnetwork}

Both the node-augmented visualisation and the ranked sociogram provide some additional information on the distribution of edges into layers, but this does not make the underlying structural representation more clear. For example, in Figure~\ref{Gr-multiplex pies} we can still see \red{several} unorganised edges in the background.

%moves away from the classical approach of graph layout in order to use the spatial position of the ranked nodes to convey meaningful information about the nodes and their relations. In this way with a single visualisation it is possible to have glimpse of the complex structure of multilayer networks receiving information about the nodes (ranking), their connections (edges), their presence on the various layers (edges colours), nodes assortativity (length of the edges) and a specific metric distribution on the whole network. We must admit the while this approach is undoubtedly able to convey a large amount of information the general look is not as obvious and clear as a traditional sociogram. On the one side sociogram have a longer history and we are used to see them in a thousands of different contexts and, on the other side, the simple geometrical metaphor of the sociograms makes almost useless any explanation. 

An opposite way is to use analytical measures to \emph{simplify} the network visualisation instead of augmenting it. Going back to Figure~\ref{Gr-multiplex pies} we can see that all the layers are entirely \red{included in the visualisation}. Here we instead propose to include only local parts of each layer. \red{For example, a}ssume that for a specific user only two layers are \emph{relevant}, e.g., they are the ones used to reach most of its neighbours, or the only way to reach some of them, or the ones defining its affiliation to some community. We can then say that locally around that user those layers are relevant and the others are \red{mainly} generating noise. Somewhere else there can be users with a different local view, for which other layers are relevant and the others represent noise.

Having a way to quantify what \emph{relevant} means, we can thus remove users (and their connections) from layers that are not relevant for them and create a new filtered multiplex providing a combination of the local views of its users. In practice, we \red{do not} ask if two layers should or should not be merged, we do not try to measure if \red{they} are \red{\emph{globally}}  similar or not, but we ask \red{these} questions \red{\emph{locally}, for each pair of nodes,} acknowledging that the answer can change when we focus on different portions of the network.

Acknowledging this aspect we introduce sociograms based on \emph{local merging} techniques. Locally merged sociograms maintain the classical look of sociograms, as well as their intuitive nature, but redefine the edges connecting the nodes.  
%In a multilayer network we might want to be able to visualise the network structure of a specific structure of the network that might exist only when several layers are observed in their intertwined nature. 
What defines an edge in this filtered sociogram is not the existence of a relationship between node \emph{a} and node \emph{b} (which is more and more probable as long as we add layers) but the existence of a connection between nodes \red{playing} the same important role for both of them. Focusing on specific multilayer metrics such as those defined in \cite{Berlingerio2012} makes it possible to visually isolate relevant network structures hidden inside the \red{multiplex network}. 

\begin{figure}
\begin{subfigure}[b]{.45\textwidth}
\centering\includegraphics[width=\columnwidth]{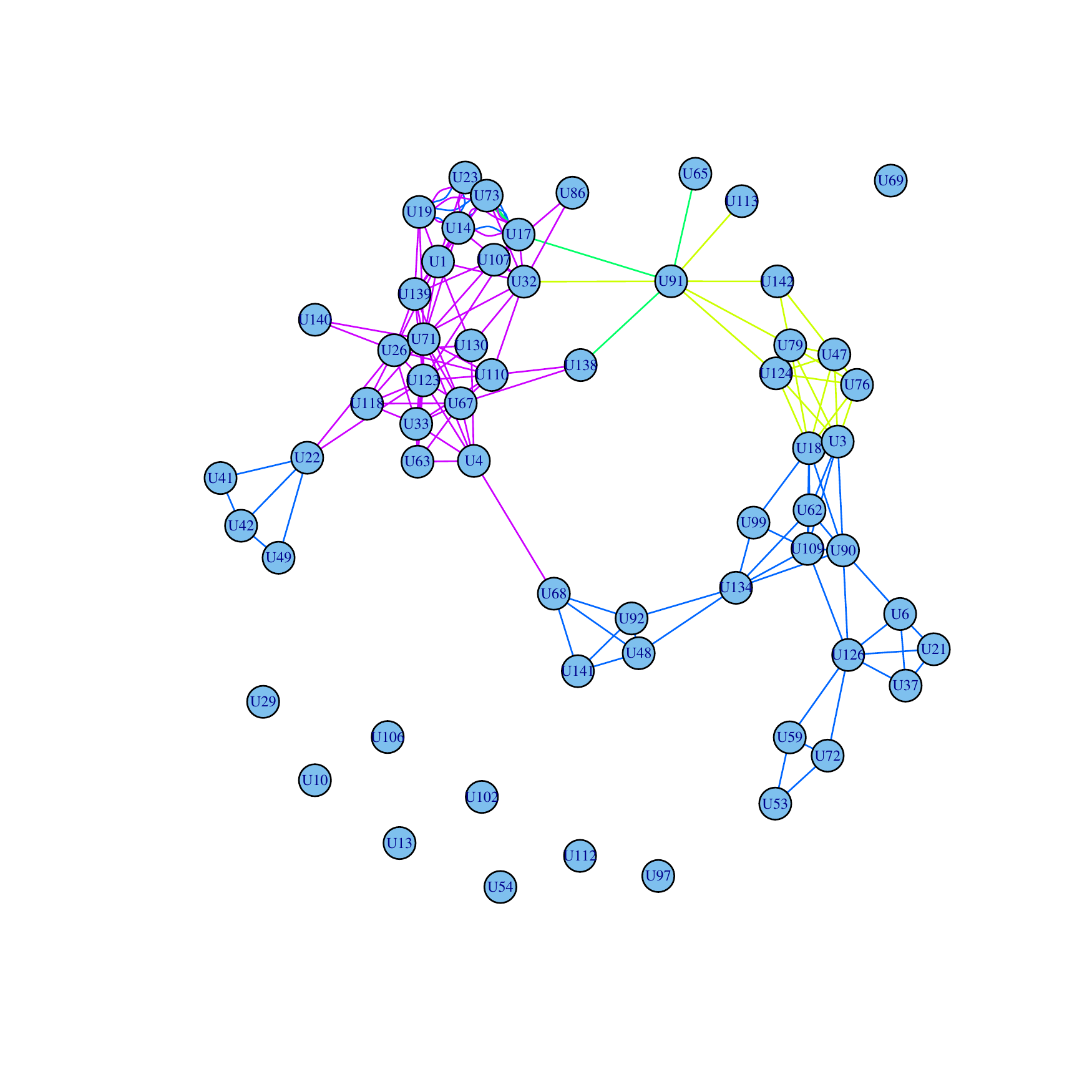}
\caption{relevance $\geq$ .6}\label{Gr-local merging relevance}
\end{subfigure}%
\begin{subfigure}[b]{.45\textwidth}
\centering\includegraphics[width=\columnwidth]{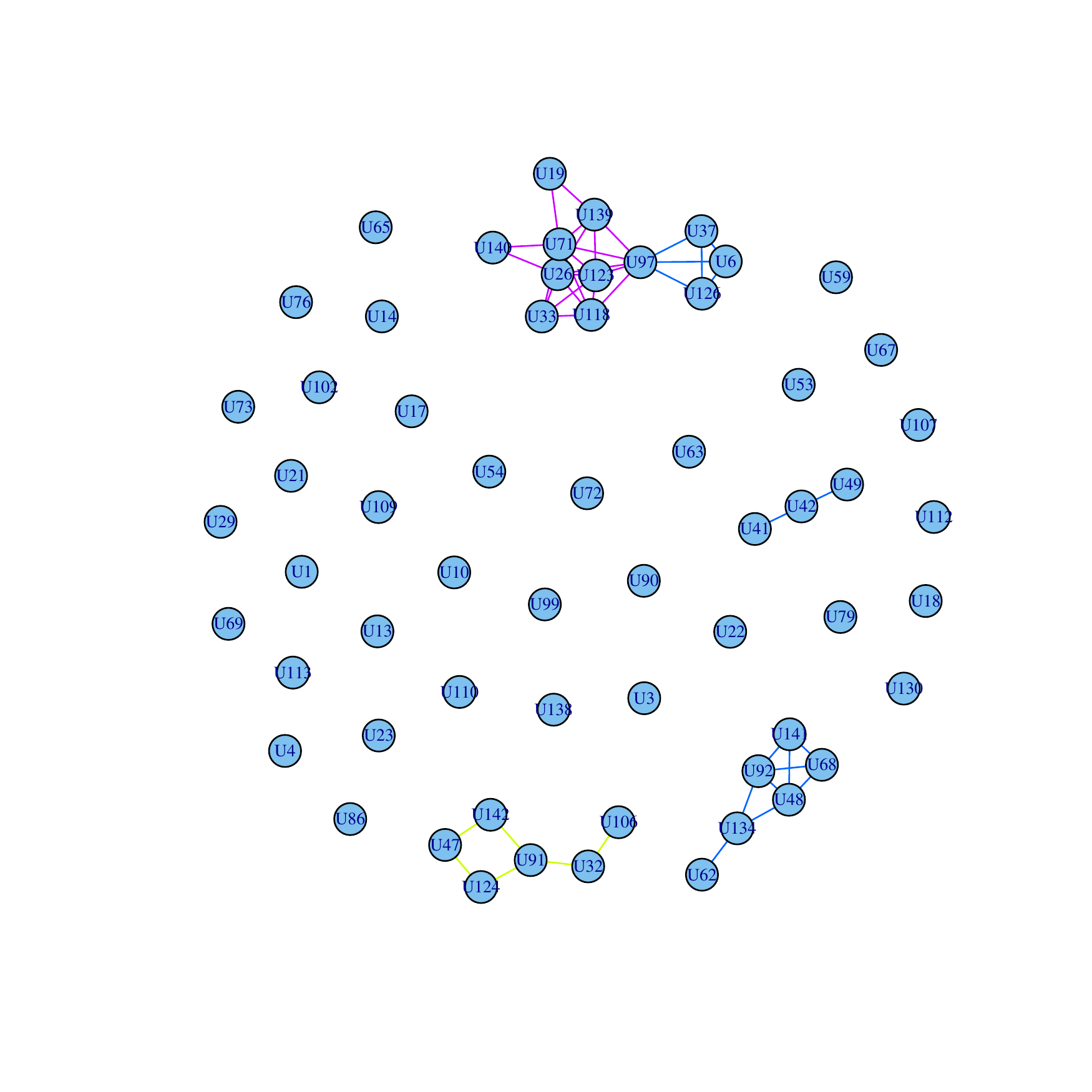}
\caption{exclusive relevance $\geq$ .3}\label{Gr-local merging xrelevance}
\end{subfigure}%
\caption{Local merging}
\label{local_merge}
\end{figure}

%In this article we present two visualisations based on the network relevance and on the network exclusive relevance. Figures \ref{Gr-local merging relevance}
%\ref{Gr-local merging xrelevance} show the two networks.  
Figure \ref{Gr-local merging relevance} shows \red{a} local merging sociogram defined according to \red{network/layer} relevance with a threshold of 0.6. 
Relevance \cite{Berlingerio2012} computes the ratio between the \red{number of} neighbours of a node on layer $d$ \red{over} the total number of its neighbours \red{on all layers}. 
%This metric deals with the fact that in a multilayer network nodes might participate in different ways on different network layers therefore their presence (and their connections) might be different when we observe different layers. Focusing from a nodes' perspective the \emph{network relevance} define how relevant is a specific layer for every user. 
Defining a local merging sociogram based on this value corresponds to selecting an edge between two nodes of the multilayer network only when a specific layer $d$ is more relevant than the input value for both of them. \red{As an example, t}he network visualised in Figure \ref{Gr-local merging relevance} shows all the existing edges connecting two nodes on a given layer that is more relevant than 0.6. 
%Comparing Figure \ref{Gr-local merging relevance} with \ref{Gr-multiplex color} allows us to make some interesting analysis. First it should be noticed that while the whole network appears to be a single component the $relevant$ network show many isolated nodes. At the same time it is worth notice that what connects together the large component of the $relevant$ network is the connection of the various layers since every user - or rather every group of users - seem to have one  or more preferred networks. While $work$ layer is, as expected due to the nature of the dataset relevant to the largest group of nodes it is interesting to notice how many users are connected only through the relevant network of $lunch$ of $Facebook$. Redefining the whole multilayer structure according to this local merging approach allows us to identify the interconnected game of the various layers without the overabundant data of the whole multilayer network.
It is important to keep in mind that this kind of visualisation considers the relevance value for the dyad and not for the single node, therefore the edges are represented only if they belong to a layer that is relevant for both the nodes of the dyad; this is why we \red{call} this a $local$ merging. 

Figure \ref{Gr-local merging xrelevance} introduces \red{a} local merging network based on exclusive relevance. Exclusive relevance \cite{Berlingerio2012} indicates the fraction of neighbours \red{of a node}  that are reachable  \red{in one step} \emph{only} through a specific layer $d$\red{, making} that specific layer essential to ensure the full connectivity of \red{the} node. While in a multilayer network nodes and connections are easily replicated through several layers\red{,} exclusive relevance measures those connections that are available only on a \red{specific subset of the layers, e.g., on a} single layer. From a node-level perspective we can assume that within a multilayer network a layer that shows a high level of exclusive relevance is \emph{used} by the node in a different way from the other layers. Nodes might have many reasons to maintain different connections in different layers and to \red{keep them} separated. \red{Extracting} the local merging \red{of a}  network based on the exclusive relevance \red{of its layers} allows us to visualise this \red{aspect} focusing on the dyads. As in the previous example, \red{in Figure \ref{Gr-local merging xrelevance} only the edges belonging to a layer with an} exclusive relevance higher than a given threshold (0.3 in this case) for both the nodes of the dyads are visualised. The \red{visualised edges are thus}  only those connecting nodes that are both using that specific layer in a \emph{different} way\red{, in particular, with} connections that are nor replicated in any other layer.

\begin{figure}
\begin{subfigure}[b]{.45\textwidth}
\centering\includegraphics[width=\columnwidth]{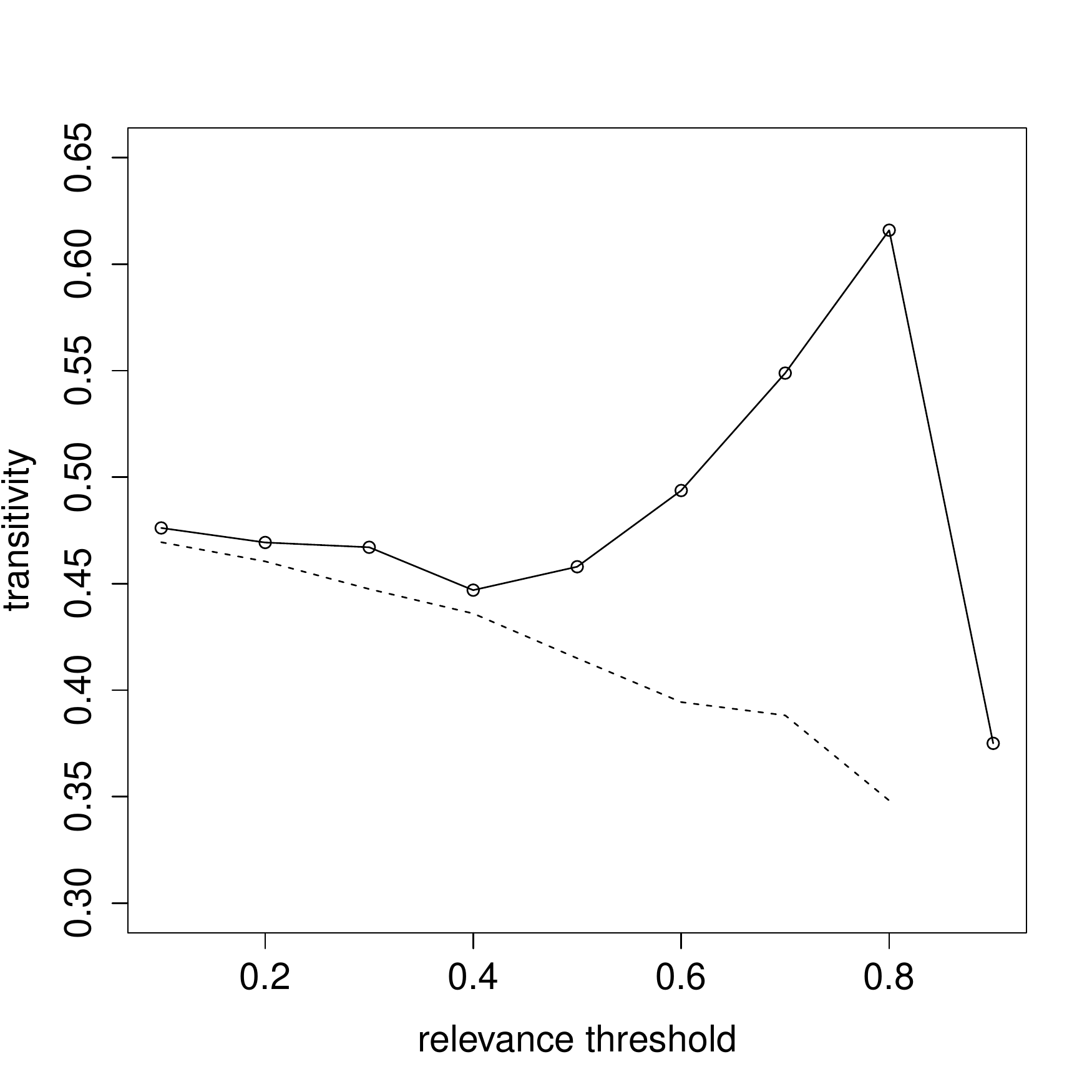}
\caption{Transitivity varying the relevance filtering threshold. Continuous line: real data. Dashed line: corresponding null model (average of 10 executions)}\label{Gr-relevance_transitivity}
\end{subfigure}%
\hspace{.5cm}
\begin{subfigure}[b]{.45\textwidth}
\centering\includegraphics[width=\columnwidth]{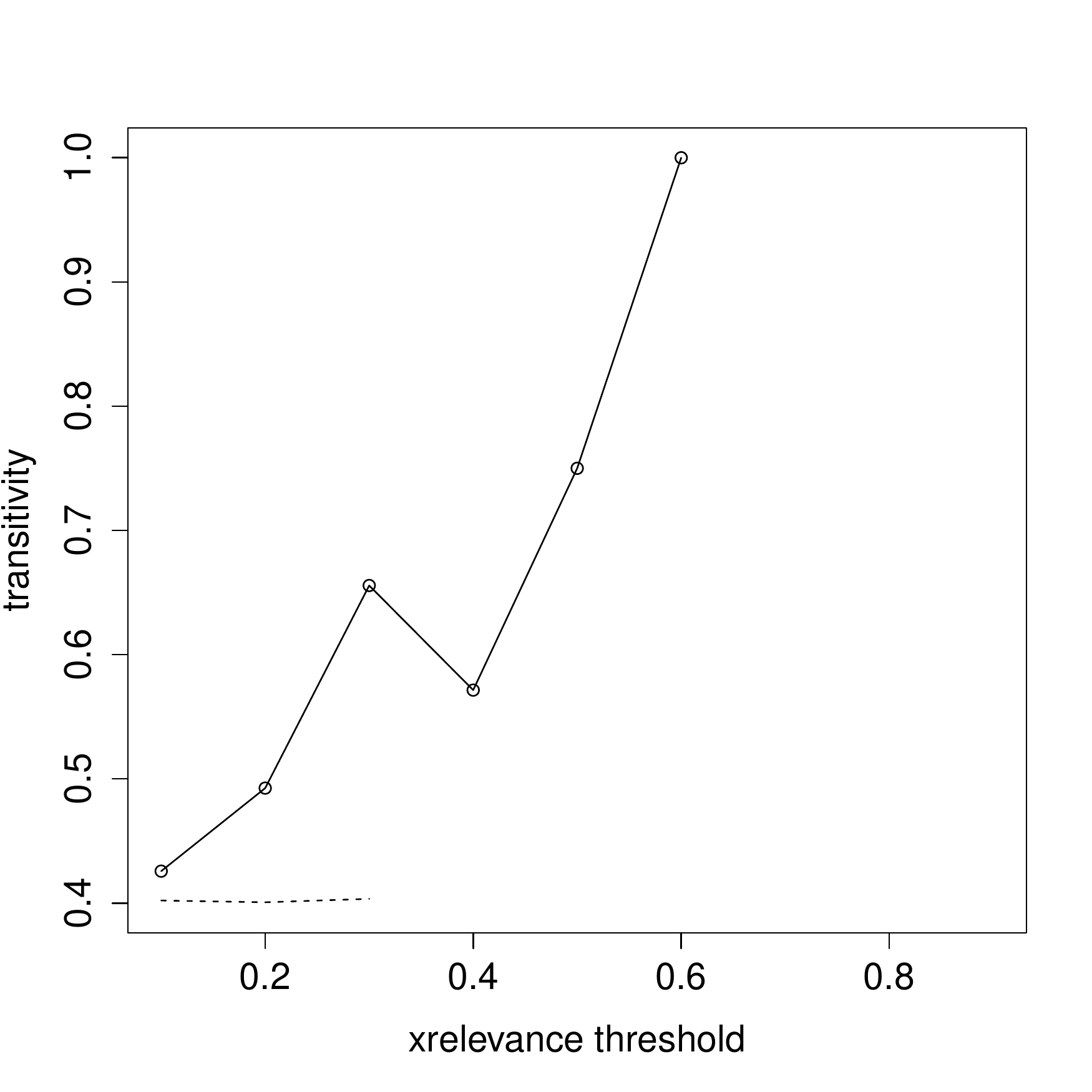}
\caption{Transitivity varying the exclusive relevance filtering threshold. Continuous line: real data. Dashed line: corresponding null model (average of 10 executions)}\label{Gr-xrelevance_transitivity}
\end{subfigure}%
\caption{Evaluation of local merging against a null model.}
\label{null-transitivity}
\end{figure}

%What can this specific metric and visualisation be used for? We claim that it might point out dynamics that would be otherwise undetectable. Local merging network based on the exclusive relevance is able to reveal hidden clusters of users connected through layers that are used only, or mainly, for those connections. Some more 
A qualitative description of this phenomenon can contribute to clarify it: if we look at \red{F}igure \ref{Gr-local merging xrelevance} and we examine the small clique on the bottom right corner \red{containing} users U141, U68, U48 and U92 we notice that these users \red{(1) are tightly connected} on the $lunch$ network \red{and (2) can only reach some of their neighbours through this network. For example, U68 is connected to U48 and U92 only because they have lunch together}. If we check their connections on the other layers we notice, from the analysis of \red{F}igure \ref{Gr-multiplex slices multi layout}\red{,} that they are all connected on the $work$ layer with U4\red{. U4} is also connected to the four nodes on the $lunch$ layer\red{,} even though \red{this is not visible in Figure \ref{Gr-local merging relevance} because this} layer\red{'s} relevance \red{is below the thresholds used in our examples. A}  possible interpretation \red{of Figure \ref{Gr-local merging relevance}} is that U4 is a central hub on the $work$ layer (this can easily be confirmed looking at Figure \ref{Gr-parabola}) and he/she has lunch with many collaborators\red{, acting as a bridge between different layers}. Nevertheless these collaborators get together only during lunches therefore their connections with other co-workers of U4 are necessarily happening \red{only exploiting} the $lunch$ network. 

%The analysis of the local merging network based on exclusive relevance allowed us to detect this complex dynamic understanding how two layers interact and are involved in nodes connectivity. The general claim we are making is that local merging techniques based on both relevance and exclusive relevance are visual tools able to detect hidden clusters within multilayer networks.
\red{To support our claim that local merging techniques based on both relevance and exclusive relevance are able to highlight hidden clusters within multilayer networks, we need to verify that these clusters are not random structures emerging as a result of randomly selecting node pairs from the different layers. More precisely, our hypothesis is that nodes for which a layer has a high relevance tend to be localized in specific parts of the layer, that is, they tend to form well-connected groups.}
In order to verify this hypothesis we computed the transitivity value of the networks \red{obtained through local merging (}based on both relevance and exclusive relevance\red{)} and we compared those with the value obtained from a random model. \red{The random model is obtained as follows: for each layer and (exclusive) relevance threshold, we count the number of nodes passing the threshold. However, while in the local merging model we preserve these nodes (and thus the edges between them), in the null model we randomly choose the same number of nodes, preserving the edges among them. If the probability of being connected for a group of nodes is not dependent on the fact that they all have high values of relevance for that layer, then we should not be able to observe a difference with the case where nodes are randomly chosen. To observe this difference, we use t}ransitivity, or clustering coefficient, \red{which} measures the probability that two nodes that are both connected to a third node are also connected to each other \cite{girvan2002community}.

Figure~\ref{null-transitivity} shows how the transitivity values are consistently higher in the measured data than in the random models, supporting our claim.
\red{The reason why in Figure~\ref{null-transitivity} the values of transitivity disappear above some thresholds depends on the fact that for high thresholds almost all edges are removed, the 
network becomes almost empty and it is no longer possible to identify hidden structures inside it. This is particularly evident when we use exclusive relevance, where it is not so common to observe people who are only or mainly present in one of the layers: if the filter is too selective we end up with an empty network, for which transitivity is undefined. Interestingly, this affects randomly sampled networks more significantly, supporting again the idea that nodes for which a layer has a high relevance tend to be localized in specific parts of the layer. The drop in transitivity in Figure~\ref{null-transitivity}(a), at relevance .9, can also be explained considering that the corresponding simplified network is almost (but not completely) empty.}

\section{Conclusion}
Generating effective visualisations of multiplex networks is an important task in exploratory analyses and reporting. So far, methods developed for simplex networks have \red{often been} adapted to multiplexes, using some additional information like colours and line types to differentiate layers. However, a problem with multiplex networks is that they provide more structural information than simplex networks and different layers may not be similar, making it impossible to find layouts which are appropriate for all of them. As a result, even a traditional representation of just nodes and edges may suffer from too much information condensed in a single diagram.

As a way to address this problem, we propose to use network properties to simplify the multiplex. Different measures can be used, leading to different meanings of the graphs. By comparing the results with null models we have seen that real data contains patterns highlighted by these representations and that are not a result of chance.

It is worth noticing that this approach is not necessarily restricted to visualisation. As an example, modern clustering approaches for multiplex networks try to find combinations of layers for which  community structure\red{s} emerge. While this is an interesting novel direction with respect to simplex network clustering, it may not make sense to combine two layers altogether. Some local portions might be correlated enough to be combined, some others might not.

As a final remark, the discussion presented in this work has only focused on the single layers composing the multiplex. However, the simplification measures used in our experiments can also be applied to sets of layers. This leads to interesting combinatorial problems to be investigated in the future.


\section{References}
%\bibliographystyle{elsarticle-num} 
%\bibliography{msna,sna,additional_refs}

\appendix

\section{AUCS dataset}
\label{dataset}

%In this section, we are going to introduce the real-world dataset which consists of measurements of multiple types of relations among the population and can be treated as a multilayer graph. We first describe the methodology behind data collection, and then provide some basic statistics of the collected data.
%- population and sampling:
The data used in the article was collected at the Department of Computer Science at Aarhus University among the employees. The population of the study is 61 employees (out of the total number of 142) who decided to join the survey, including professors, postdoctoral researchers, PhD students and administration staff.

%- variables that we measured:
%According to \cite{Wasserman1994}, there are two types of variables that can be included in a network data set: \emph{structural} and \emph{composition} variables. Structural variables are measured on pairs of actors and they express specific ties between the pairs (e.g. friendship). Composition variables are defined for individual actors and they are measurements of various actor attributes (e.g. age). 
%
For our study, we measured 5 structural variables, namely: current working relationships,  repeated leisure activities, regularly eating lunch together, co-authorship of a publication, and friendship on Facebook.
These variables cover different types of relations between the actors based on their \emph{interactions}. All relations are \emph{dichotomous} which means that they are either present or absent, without weights.
%TODO: if we have groups or ranks, we also have composition vars
%TODO: idea: how about adding information that we know about our group by hand??

%- collection process (offline):
Measurements of the first three variables (off-line data) were collected via a \emph{questionnaire} which had been distributed among the employees on-line. 
%The questions were of a \emph{roster} format which means that each actor was presented with a complete list of other actors in the network and asked to select people with whom he/she has the aforementioned ties (working relationship, leisure activities, eating lunch). The number of choices for each question was not limited. 
The measurements are results of individual assessments done by the actors and therefore the relations were directed, however, we do not intend to study the aspect of personal perception of the relationships and so we decided to flatten the data into nondirectional connections. Thus, if an actor $A$ indicated a tie to actor $B$, we input an edge into the network even if actor $B$ did not indicate a tie to actor $A$.
On the top of this, the respondents were asked to provide their user information for a couple of most widespread online networks. 77\% of the respondents who filled in the questionnaire stated that they have a Facebook account and provided their username. All respondents provided answers to all questions which means that our multi-layer network is complete.
%In total, 50\% people responded to the survey and provided answers to all questions. This of course affects the accuracy of measurements, nevertheless, since we flatten all relations into nondirectional, the probability that we miss a tie is approximately 0.25. We do not claim that with such a collection we can perfectly reconstruct true relations in the population but the respondents who joined the survey provided information about all types of their ties which means that the multi-layer patterns in the network (if there are any) should be preserved.
%- collection process (online):
%TODO: is coauthorship online or offline?
Information about the co-authorship relation was obtained from the on-line DBLP bibliography database without the need to directly ask the actors. A co-authorship of at least one publication by a pair of actors resulted in an edge in the network. DBLP gets new data with a delay of several months and therefore the ``current working relationships" network is quite distinct from it. Moreover, ``current working relationships" network includes other types of interactions than cooperation on papers (e.g. also cooperation on administrative work).

Friendship relations among all the actors who stated that they have a Facebook account were retrieved from the site using a custom application.
%TODO: what else can we say?

Finally, in Table~\ref{tabDataDescription} we describe some common statistical measures of the single-layer networks. The Co-authorship network is the smallest and less connected of all layers, Work and Lunch have the most edges and the highest average vertex degree can be observed for the Facebook layer.

\begin{table}[t!] %\scriptsize
\begin{center}
\begin{tabular}{lccccc}
\hline
  &   \small{Work} &  \small{Leisure}  &  \small{Coauthor}  & \small{Lunch} &  \small{FB} \\
\hline
 \small{\# of edges}  &   194  &  88 &  21 &  193 &   124 \\
\small{\# of con. comp.} &  2   &  1   &   8 &   1  &   1    \\
 \small{avg. vertex deg.} &  $6.47$ &   $3.74$  &  $1.68$ &  $6.43$ &   $7.75$   \\ 
\hline
\end{tabular}
\caption{Basic statistics computed on the sociomatrices of the 5 different relations---number of edges, number of connected components and average vertex degree} 
\label{tabDataDescription}
\end{center}
\end{table}

\section{Visualising Multiplex network metrics}
\label{metrics}
%Expanding our approach into a multilayer network perspective introduces both new interpretation of existing metrics as well as new metrics that were impossible to measure within a single layer approach. What we present here does not aim at being an ultimate comprehensive list of every possible metric that can be visualised but it is more intended as a starting point based on a set of proposed visualisation for this kind of data.
%

\begin{figure}
\begin{subfigure}[b]{.32\textwidth}
\centering\includegraphics[width=\columnwidth]{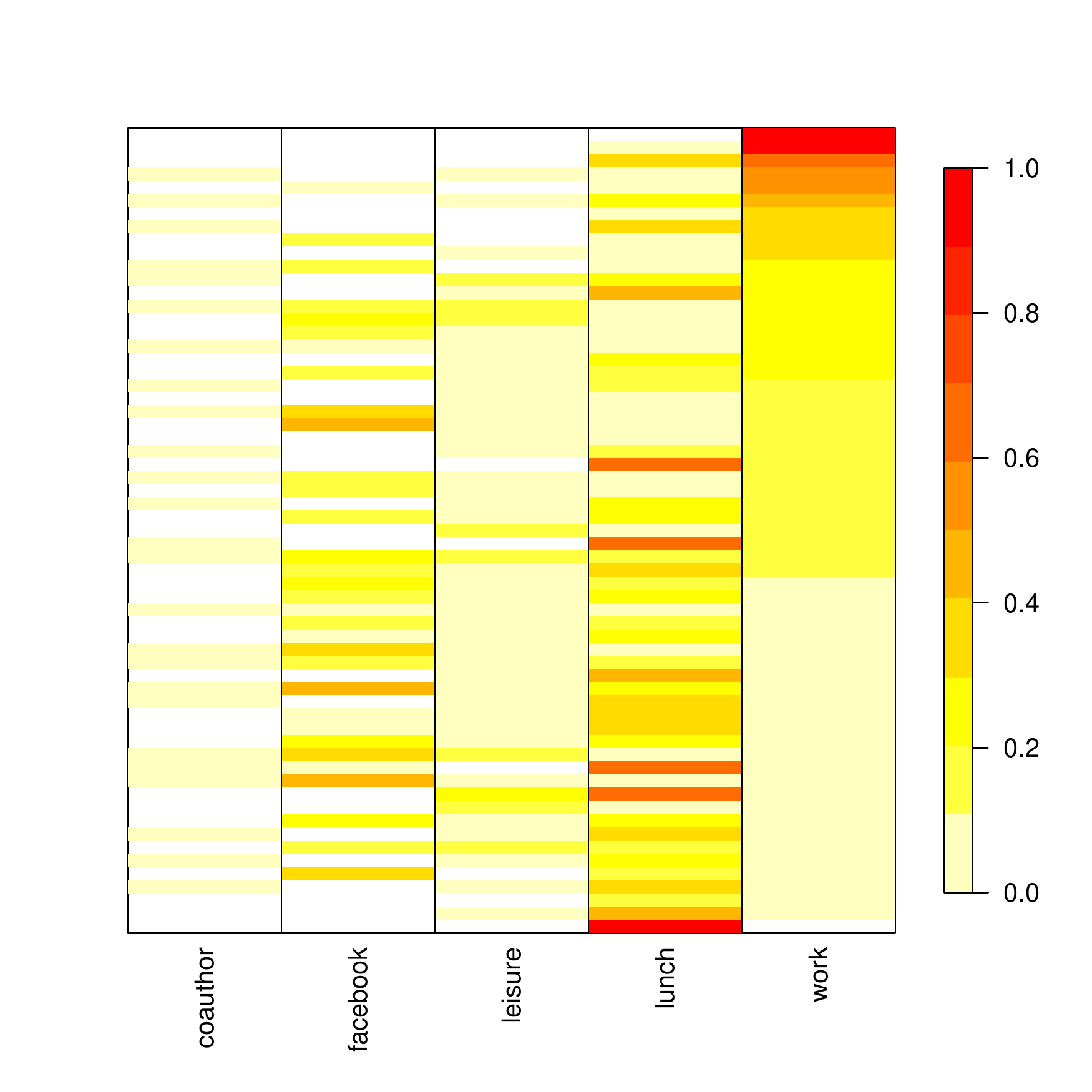}
\caption{}\label{Gr-par1}
\end{subfigure}%
\begin{subfigure}[b]{.32\textwidth}
\centering\includegraphics[width=\columnwidth]{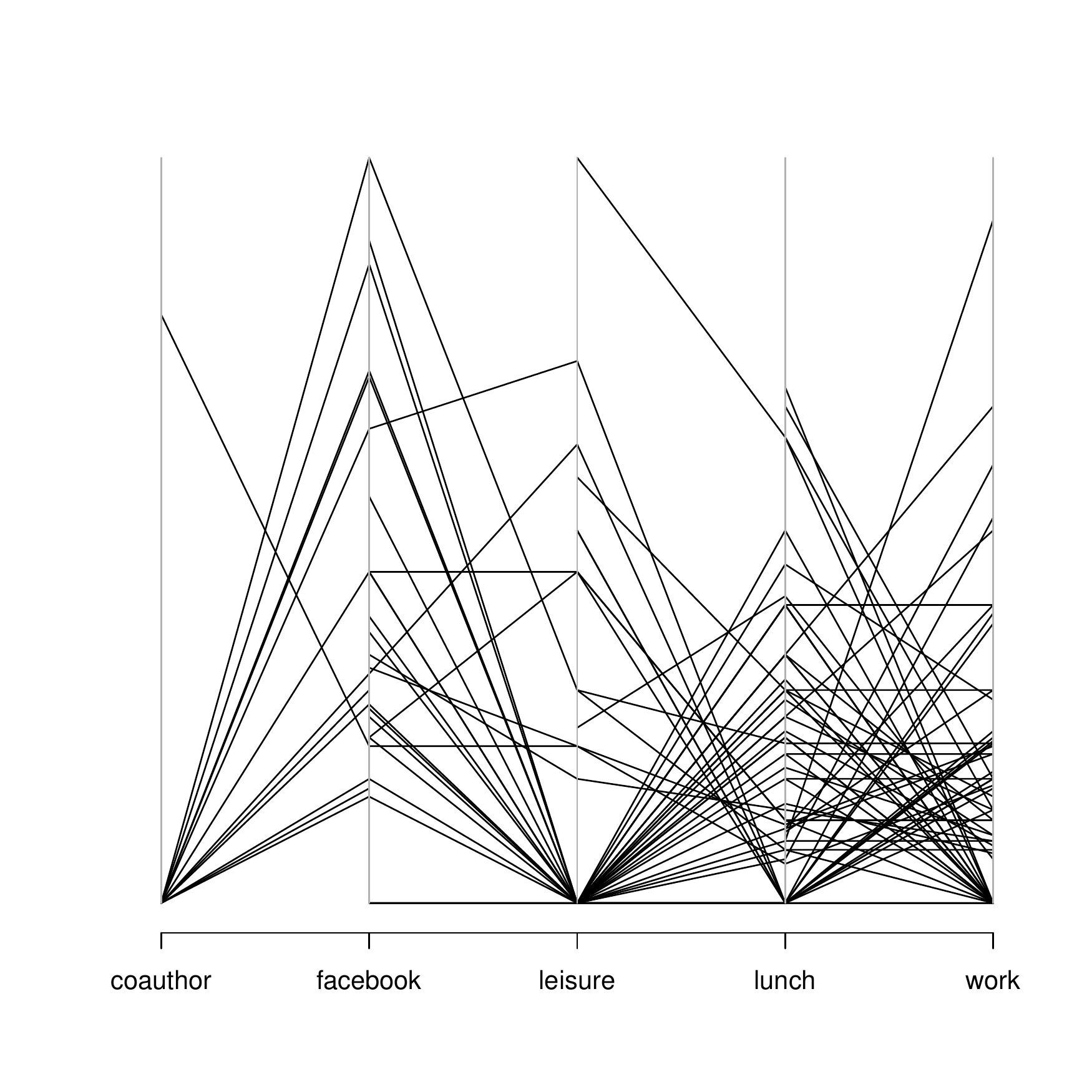}
\caption{}\label{Gr-par2}
\end{subfigure}%
\begin{subfigure}[b]{.32\textwidth}
\centering\includegraphics[width=\columnwidth]{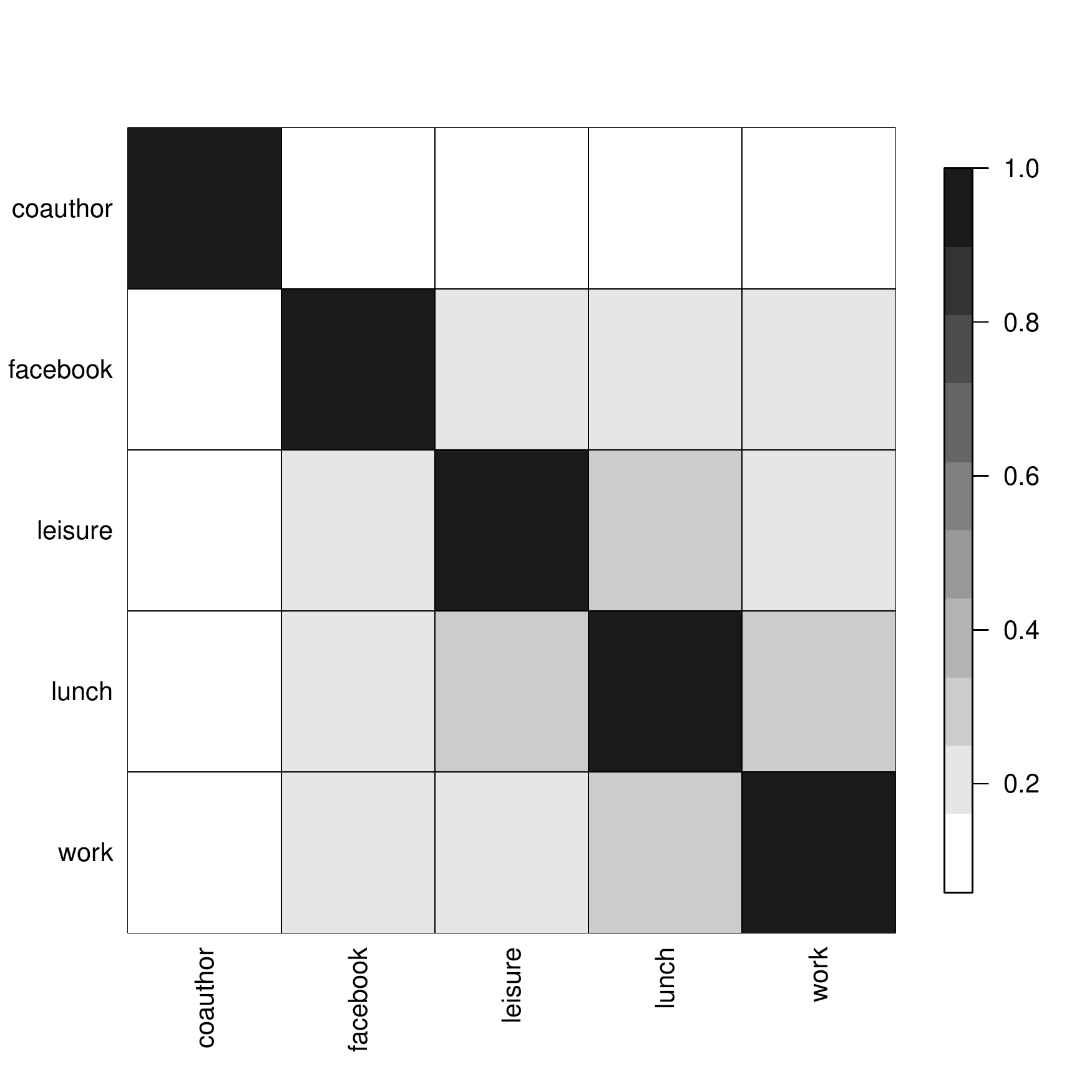}
\caption{}\label{Gr-netsim}
\end{subfigure}%
\caption{Metrics}
\label{m:metrics}
\end{figure}

%\subsection{Parallel coordinates visualisation}
Multiplex networks analysis metrics can be measured on every single layer or combination of layers. As a consequence every node can behave differently on different layers, e.g., a user might be highly connected on Facebook and have almost no followers on Twitter. 
%This difference of behaviour is an additional relevant information about that user. Information here does not generate only from the measurement of a specific metric on a specific layer but also from how the values measured on different layers relate to each other. 
Parallel coordinates are typical visualisations for multi-dimensional data \cite{moustafa2011parallel}.
Figures \ref{Gr-par1} and \ref{Gr-par2} use two variations of a parallel coordinates plot to visualise some of the metrics defined by  \cite {matteoMDMea2013}. 
%Figure \ref{Gr-neighbourhood} show the number of neighbours for every user on every layer of the multiplex network. The value of neighbourhood is defined as the number of nodes directly connected with one node the a specific layer. It is interesting to notice that the final shape designed by the line in the parallel coordinates visualisation represent a specific profile of how a single user is connected trough the various layer and that the shape of the line will be the same despite of the number of connections.
%Traditional limitations of this visualisation are still valid: order of the parallel axes is arbitrary and must be clearly showed, at the same time a common scale is required. The latter is a minor concern since the visualisation is supposed, in this specific context, to show the same metric using the same scale on every single layer. Figure \ref{Gr-relevance} shows the  value of relevance\cite {matteoMDMea2013} for every user in every layer of the network. The relevance of a layer measures the ratio between the neighbours of a node on a specific layer and the total number of its neighbours. It should be noticed how in this case the parallel coordinates visualisation would not provide any useful information due to the nature of the metric we are visualising. The values that the relevance might assume on the various layers are, in fact, inversely proportional therefor the shape defined by the data will be largely due to the nature of the data itself.

%\subsection{Block matrix visualisation for layer correlation}
Multiplex network analysis also introduces the possibility to investigate the correlation between two pairs of layers belonging to the same network. A block matrix visualisation can be used to visualise this kind of information. Figure \ref{Gr-netsim} shows the values of the Jaccard network correlation   
\cite{Berlingerio2012} for the various pairs of layers of the AUCS multilayer network. While in this case interpretation is straightforward it is still interesting to notice how following a similar approach it is possible to analyse existing correlations not only between pairs of layers but also between any pair of sets of layers.

\section{More on the ranked sociogram}
\label{ranked}
\begin{figure}
%\begin{subfigure}[b]{.5\textwidth}
%\centering\includegraphics[width=\columnwidth]{img/ba_corr}
%\caption{BA networks, correlated}\label{fig:BA1}
%\end{subfigure}%
\begin{subfigure}[b]{.5\textwidth}
\centering\includegraphics[width=\columnwidth]{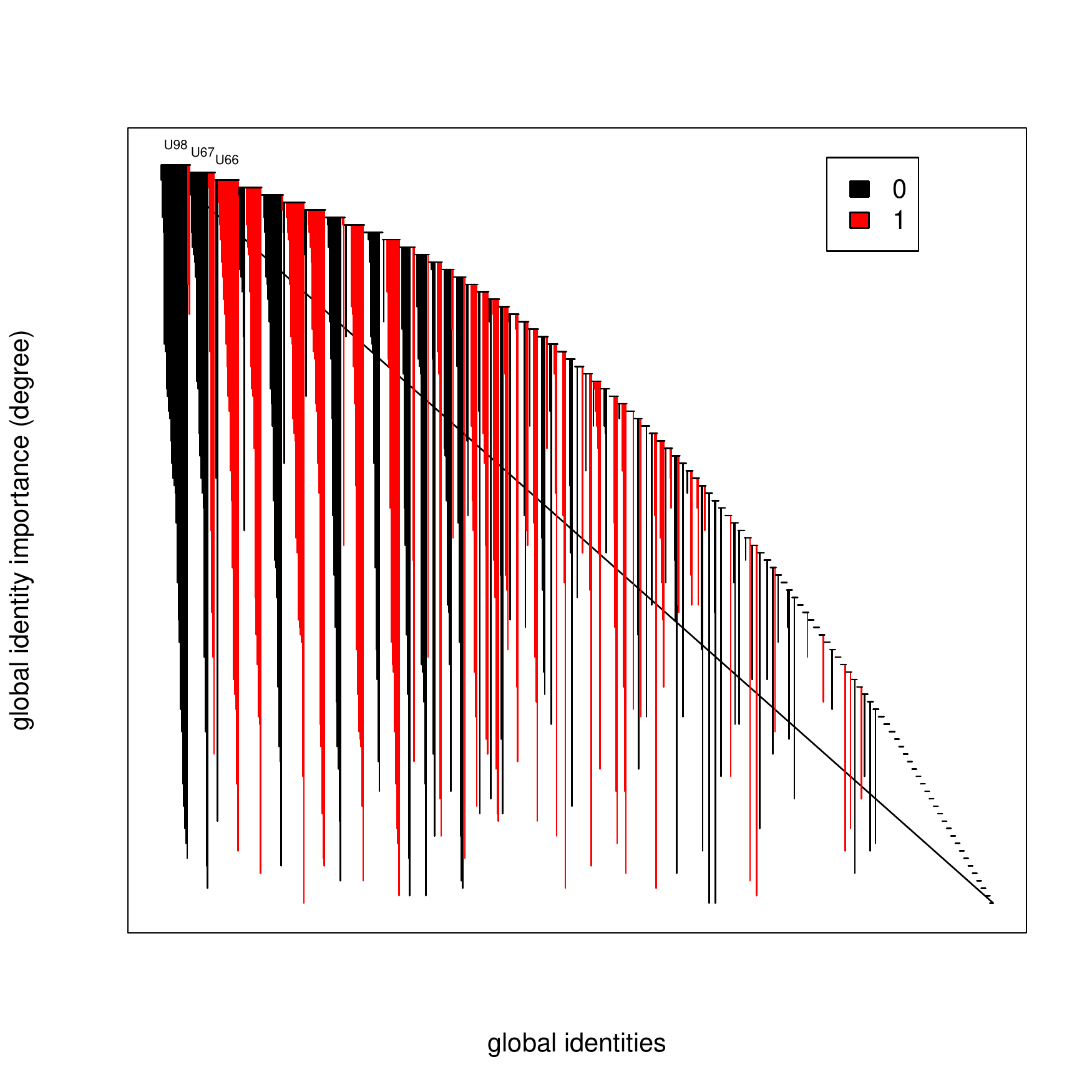}
\caption{BA networks}\label{fig:BA2}
\end{subfigure}
%\begin{subfigure}[b]{.5\textwidth}
%\centering\includegraphics[width=\columnwidth]{img/unif_corr}
%\caption{Uniform networks, correlated}\label{fig:Un1}
%\end{subfigure}%
\begin{subfigure}[b]{.5\textwidth}
\centering\includegraphics[width=\columnwidth]{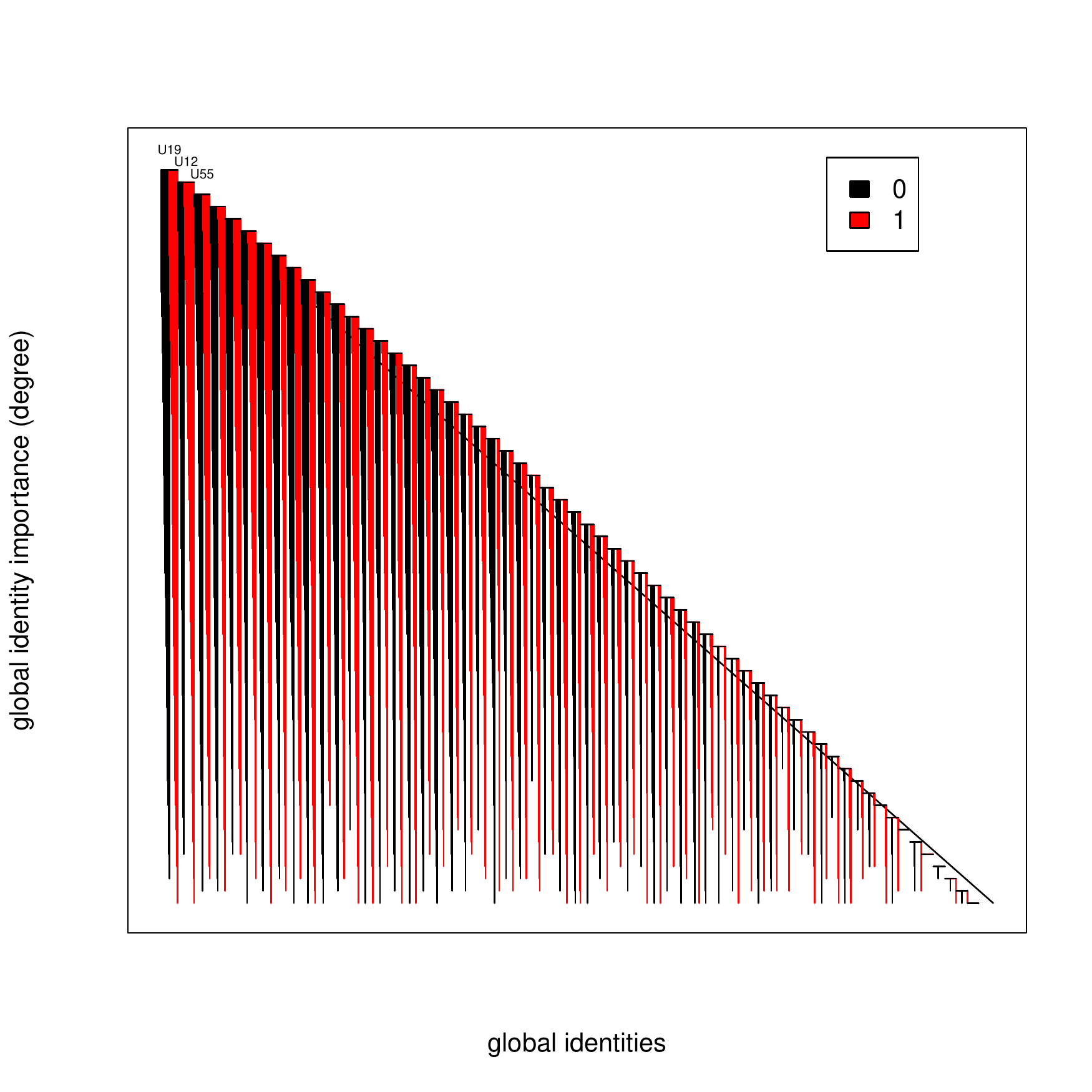}
\caption{Uniform networks}\label{fig:Un2}
\end{subfigure}%
\caption{Ranked sociogram on generated networks}
\label{ranked-generated}
\end{figure}

A ranked sociogram is able to convey network level information such as the distribution within the multi-layer structure of a specific metric  (in this case the degree distribution). 
Figure~\ref{ranked-generated} shows the ranked sociogram of the degree distribution for two multi-layer networks generated according to an extended Barabasi-Albert and an Erd\"{o}s-ReŽnyi model generated using the framework in \cite{Magnani2013} and the multinet package\footnote{https://github.com/magnanim/multiplenetwork}.
What can be noted is that in a single picture we have an overview of a large quantity of information both about the general multiplex network (degree distribution), about the nodes behaviour on the various levels (distribution of black and red connections) and about the assortativity or dissortativity of the networks (length of the edges). 

%\begin{thebibliography}{00}


%\end{thebibliography}
\end{document}